\documentclass[letter]{aa}  

\usepackage{geometry}
\usepackage{amsmath}
\usepackage{amssymb}
\usepackage{graphicx}
\usepackage{txfonts}
\usepackage{natbib}
\bibpunct{(}{)}{;}{a}{}{,} 

\newcommand\cm{\,\rm cm}

\newcommand\erg{\,\rm erg}
\newcommand\K{\,\rm K}

\newcommand\Myr{\,\rm Myr}
\newcommand\Gyr{\,\rm Gyr}

\newcommand\kms{\,\rm km\,s^{-1}}
\newcommand\pc{\,\rm pc}
\newcommand\kpc{\,\rm kpc}
\newcommand\cms{\,\rm cm^{2}s^{-1}}

\newcommand\tms{\!\times\!}

\newcommand\yy{\hat{{\mathbf y}}}

\newcommand{\mn}[1]{\overline{#1}}
\newcommand{\rms}[1]{\left<\right.\!#1\!\left.\right>}

\newcommand\U{\mathbf{u}}
\newcommand\mU{\mn{\mathbf{u}}}
\newcommand\B{\mathbf{B}}
\newcommand\mB{\mn{\mathbf{B}}}
\newcommand\EMF{\mathcal{E}}

\begin{document}

  \title{Direct simulations of a supernova-driven galactic dynamo}
  \titlerunning{SN-driven galactic dynamo}

  \author{Oliver Gressel, Detlef Elstner, Udo Ziegler, 
          and G{\"u}nther R{\"u}diger }
  \authorrunning{O. Gressel, D. Elstner, U. Ziegler, \& G. R{\"u}diger}

  \offprints{O. Gressel}
  \institute{Astrophysikalisches Institut Potsdam, 
             An der Sternwarte 16, D-14482 Potsdam, Germany\\
             \email{ogressel@aip.de}}
   \date{}

  \abstract
  { Supernovae are known to be the dominant energy source for driving
    turbulence in the interstellar medium. Yet, their effect on magnetic field
    amplification in spiral galaxies is still poorly understood. Previous
    analytical models, based on the evolution of isolated, non-interacting
    supernova remnants, predicted a dominant vertical pumping that would render
    dynamo action improbable.}
  { In the present work, we address the issue of vertical transport, which is
    thought to be the key process that inhibits dynamo action in the galactic
    context. We aim to demonstrate that supernova driving is a powerful
    mechanism to amplify galactic magnetic fields.}
  { We conduct direct numerical simulations in the framework of resistive
    magnetohydrodynamics. Our local box model of the interstellar medium
    comprises optically-thin radiative cooling, an external gravitational
    potential, and background shear. Dynamo coefficients for mean-field models
    are measured by means of passive test fields.}
  { Our simulations show that supernova-driven turbulence in conjunction with
    shear leads to an exponential amplification of the mean magnetic field.
    We found turbulent pumping to be directed inward and approximately
    balanced by a galactic wind. }
  {}


   \maketitle


\section{Introduction}\label{intro}

Large scale magnetic fields, as observed in disk galaxies, are thought to be
the result of a turbulent dynamo. Mean-field models based on the turbulent
$\alpha$-effect can indeed explain most of the observed features in many
nearby disk galaxies \citep[cf.][]{1996ARA&A..34..155B}.  However,
uncertainties remain in the estimation of the turbulent transport coefficients
and the nonlinear saturation process. Additionally, the helicity conservation
in ideal magnetohydrodynamics (MHD) could lead to the so-called catastrophic
quenching \citep{2005PhR...417....1B}, a saturation process that scales with
the magnetic Reynolds number. Helicity transport by a wind, as considered by
\citet{2007MNRAS.377..874S}, would be one solution to this scenario.

Recently, corrobative evidence has been found that the degree of regularity of
magnetic fields decreases with increasing star formation activity
\citep{2007arXiv0712.4175C}. This has renewed the theoretical interest in
dynamo models based on the interaction of supernova-driven turbulence with
galactic differential rotation. 

Despite the success of mean-field models, until now there has been no clear
numerical verification of the dynamo process in the galactic context, and
semi-analytical models, which are based on doubtful assumptions, only have 
limited predictive power \citep[see][for a recent account]{2008AN....329..619G}.

\cite{2004ApJ...617..339B} found growing small scale fields for a non-rotating
isothermal gas. Similarly, \cite{2005A&A...436..585D} considered the most
realistic model of the interstellar medium (ISM) to date, but did not include
the rotation and shear necessary for a mean-field galactic dynamo to operate.
Except for their neglect of thermal instability (TI), the simulations of
\cite{1999ApJ...514L..99K} are very similar to ours. However, due to the
limited computational resources available at the time, their setup suffered
from a too small box size, which prohibited a long-term evolution into
developed turbulence.

\cite{1992ApJ...401..137P} proposed the magnetic buoyancy instability
supported by cosmic rays as a driving mechanism for the galactic dynamo.
\cite{2004ApJ...605L..33H,2006AN....327..469H} performed simulations of this
process, employing the injection of cosmic ray energy into an adiabatic gas as
the driving mechanism. In these simulations, cosmic ray transport is treated in
the (anisotropic) diffusion approximation. The authors found an amplification
of the mean magnetic field with an e-folding time on the order of $100\Myr$.
Contrary to these cosmic ray models we consider the injection of thermal
energy of clustered supernovae and apply a radiative cooling function to
reflect the multi-phase nature of the ISM. Although currently one cannot yet
resolve all the complex effects of the interstellar plasma, going beyond
simple artificial forcing (relying on ad-hoc assumptions), these new
approaches mark the first step towards realistic dynamo models based on
fundamental physics.


\section{Methods}\label{methods}

We explore the evolution of the galactic magnetic field within the stratified,
turbulent ISM by means of combined 3D, resistive MHD simulations employing the
NIRVANA code \citep{2004JCoPh.196..393Z} and 1D mean-field models based on
turbulence parameters.

\subsection{Direct simulations}

The adopted domain covers a vertically centred
box of $0.8\times 0.8\times 4.267\kpc^3$, representing a local patch of the
galactic disk. The physical model includes an external gravitational potential
\citep{1989MNRAS.239..605K}, differential rotation in the shearing box
approximation \citep{2007CoPhC.176..652G}, and optically-thin radiative
cooling. For the latter, we apply a simple piecewise power-law description
\citep{2002ApJ...577..768S,2005MNRAS.356..737S}. In the following, we will
briefly introduce the key parameters of our simulations. For a more detailed
description of the model, we refer the reader to \cite{2008AN....329..619G}.

We choose an initial midplane density of $n_0=1\cm^{-3}$ at a mean molecular
weight of $\bar{\mu}=0.6$. This sets the equilibrium midplane pressure at a
value of $p_0/k_{\rm B}=6000\K\cm^{-3}$. The initial stratification is an
exact hydrostatic equilibrium solution in the absence of the kinetic pressure
caused by the SNe. Contrary to previous isothermal solutions, our approach
applies an effective equation of state given by the additional constraint that
the initial disk is in equilibrium with respect to the radiative cooling. The
weak initial magnetic field has a radial and azimuthal component with
${B_{\phi}/B_R=-10}$. The magnitude of the field as a function of height is
scaled with $(p(z)/p_0)^{1/2}$ to yield a constant plasma parameter of
$\beta_{\rm P}=2\tms10^7$ throughout the disk. The rotation period is measured
in units of ${\Omega_0=25\kms\kpc^{-1}}$, while differential rotation is
quantified via the shear parameter $q={\rm d}\ln\Omega/{\rm d}\ln R$. We apply
a constant kinematic viscosity of $\nu=5\tms10^{24}\,\cms$ along with a
microscopic magnetic diffusivity of $\eta=2\tms10^{24}\,\cms$, resulting in a
magnetic Prandtl number of ${\rm Pm}=2.5$. For reference, we also define the
magnetic Reynolds number ${\rm Rm}=L_x^2\Omega_0/\eta \simeq 2500$.

Turbulence is driven solely via the localised injection of thermal energy.
While type-I SNe are exponentially distributed with a scale height of
$325\pc$, for the more frequent type-II SNe, we adopt a prescription where the
probability distribution is determined by the vertical gas density profile.
This avoids an artificial fragmentation of the inner disk observed in
conjunction with a static SN distribution. The reference explosion frequencies
are $\sigma_0^{\rm I}=4\Myr^{-1}\kpc^{-2}$ and $\sigma_0^{\rm
  II}=30\Myr^{-1}\kpc^{-2}$, respectively, and the associated energies are
$10^{51}$ and $1.14\tms10^{51}\erg$, where the higher value for the latter
accounts for the contribution from a stellar wind of the massive progenitor
\citep{2001RvMP...73.1031F}.

\subsection{Mean-field model}

We follow the common paradigm where the magnetic field $\B$ and the velocity
$\U$ are separated into mean part $\mB$ and $\mU$ plus fluctuations denoted by
$\B'$ and $\U'$. In this framework, the amplification of the mean field is
described via an averaged induction equation that comprises an additional term
$\EMF$ subsuming the action of the turbulent flow, i.e.,
\begin{equation}
  \partial_{t}\mB = \nabla \times 
                    \left[\ \left(\mU+q\Omega x\,\yy\right)\tms\mB 
                         + \EMF - \eta \nabla\tms\mB\ \right]\,,
  \label{eq:mf_ind}
\end{equation}
where $\eta$ is the microscopic magnetic diffusivity and the term including
$q\Omega$ describes the induction due to the background shear.

From the direct simulations described above, we compute the turbulent mean
electromotive force $\EMF = \mn{\U'\tms\B'}$ by applying horizontal
averages. From the obtained vertical profiles $\EMF(z,t)$ and $\mn{u}_z(z,t)$
we successfully reconstruct the mean magnetic field $\mB(z,t)$ via 
Equation~(\ref{eq:mf_ind}) demonstrating the applicability of the chosen 
mean-field approach.

As a simple closure for the mean induction equation, that via $\EMF$ still
depends on the small-scale flow $\U'$ and magnetic field $\B'$, we assume a
standard parameterisation of the form
\begin{equation}
  \EMF_i = \alpha_{ij} \bar{B}_j 
         - \tilde{\eta}_{ij}\varepsilon_{jkl}\partial_k \bar{B}_l\,,
  \quad i,j \in \left\{R,\phi\right\}, k=z\,,
  \label{eq:param}
\end{equation}
with tensorial parameters $\alpha_{ij}$ and $\tilde{\eta}_{ij}$. We apply 
the test-field method of \citet{2005AN....326..245S,2007GApFD.101...81S} 
to invert this tensor equation for given test fields and their associated
EMFs. In particular, we use the fields proposed by \citet{2005AN....326..787B}
for the shearing box case 
\citep[see also][]{2008ApJ...676..740B,2008AN....329..619G}.

While the diagonal elements of ${\boldsymbol \alpha}$ produce the dynamo
process itself, its antisymmetric off-diagonal part is responsible for the
vertical pumping $\gamma_z=\frac{1}{2}(\alpha_{\phi R}-\alpha_{R\phi})$.
Similarly, the diagonal elements of ${\tilde{\boldsymbol\eta}}$ are
interpreted as turbulent resistivity $\eta_{\rm
  t}=\frac{1}{2}(\tilde{\eta}_{RR}+\tilde{\eta}_{\phi\phi})$, while its
antisymmetric off-diagonal components can lead to
${\boldsymbol\Omega}\tms{\mathbf J}$-type dynamo effects
\citep{1969MDAWB..11..272R}, determined by
$\delta_z=\frac{1}{2}(\tilde{\eta}_{R\phi}-\tilde{\eta}_{\phi R})$.


\section{Results}\label{results}

\begin{figure}\begin{center}
  \includegraphics[width=0.90\columnwidth]{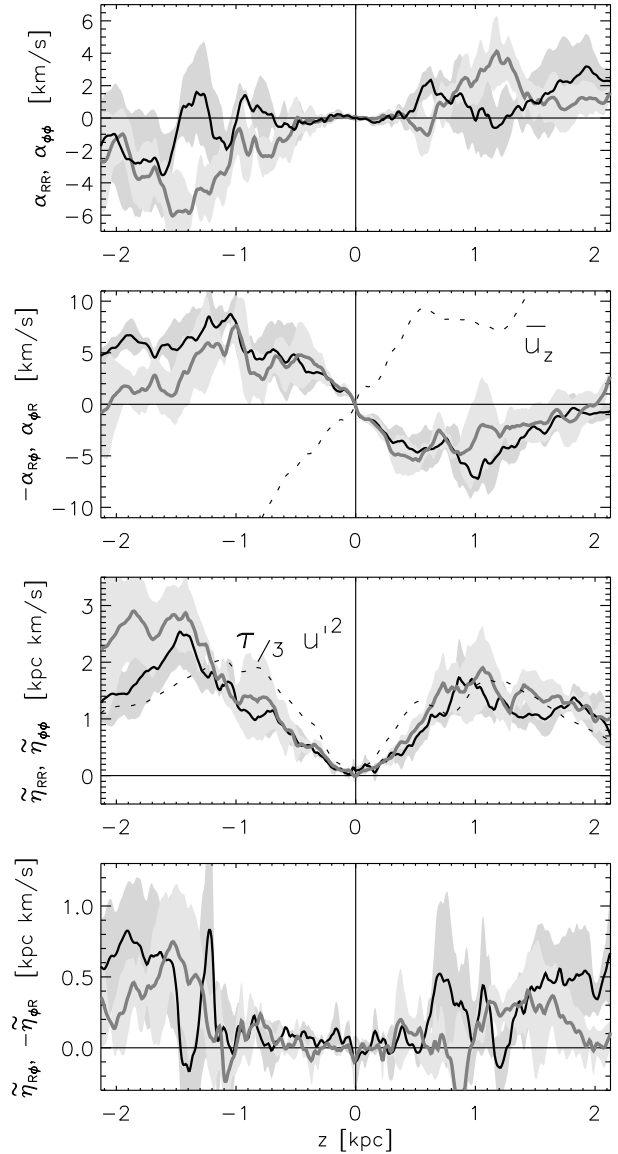}
  \caption{Dynamo coefficients for model H4. Starting at $t=1.3\Gyr$ the 
    TF-method is applied to four consecutive time intervals of $25\Myr$ each. 
    Quantities indicated by the ordinate labels are plotted in dark
    ($\alpha_{RR},\dots$) or light ($\alpha_{\phi\phi},\dots$)
    colours, respectively. Shaded areas indicate $1\sigma$-fluctuations.}
  \label{fig:alpha_eta}
\end{center}\end{figure}

We here compare six models (see Table~\ref{tab:models}) with varying SN-rate
$\sigma$ and rotation frequency $\Omega$, keeping the shear parameter $q$ and
the ratio $\sigma^{\rm I}:\sigma^{\rm II}$ fixed. From the initial
configuration turbulence builds up smoothly within about $50\Myr$ and the disk
reaches a quasi-stationary state. The vertical stratification is now
determined by the additional kinetic pressure from the SNe.

\begin{figure*}\begin{center}
  \includegraphics[width=1.90\columnwidth]{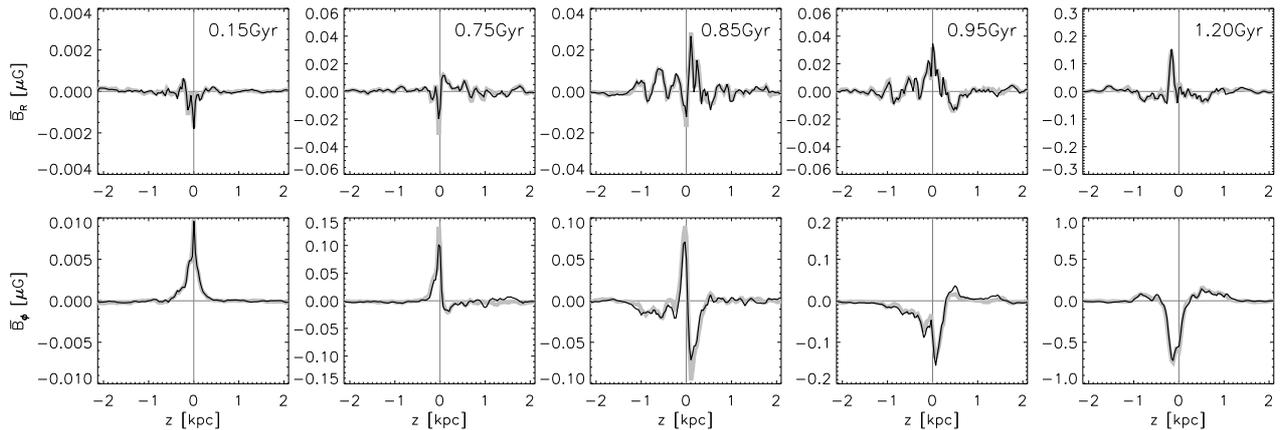}
  \caption{Profiles of the regular radial and azimuthal field for model H4 
    at various times. Results of the simulation (grey lines) are compared with
    the fields computed from $\EMF(z,t)$ via eq~(\ref{eq:mf_ind})
    (black lines). At $t\approx 0.85\Gyr$, a field reversal with pronounced 
  dipolar symmetry occurs.}
  \label{fig:B_of_z}
\end{center}\end{figure*}

\begin{table}\begin{center}
  \begin{tabular}{lcccccc} \hline
  & Q4 & H4 & F4 & F1 & F2 & F8 \\\hline
  $\sigma/\sigma_0$ & 0.25 & 0.50 & 1.00 & 1.00 & 1.00 & 1.00 \\
  $\Omega/\Omega_0$ & 4.0  & 4.0  & 4.0  & 1.0  & 2.0  & 8.0  \\\hline
\end{tabular}\end{center}
\caption{Overview of conducted models. The letters 'Q', 'H', and 'F' indicate 
         quarter, half, and full SN-rate while numbers give the rotation rate. 
         All models include shear with $q=-1$.
         \label{tab:models}}
\end{table}

\subsection{Dynamo parameters}

The most distinctive finding persistent in all our runs is that despite the
peaked distribution of the SNe the velocity dispersion increases with galactic
height for the inner $2\kpc$ of our box. This is also reflected in the lower
two panels of Fig.~\ref{fig:alpha_eta} where we plot the ${\boldsymbol
  \alpha}$ and ${\tilde{\boldsymbol\eta}}$ parameters obtained via the
test-field method. In excellent agreement with the prediction of second-order
theory \citep[SOCA, ][]{2004maun.book.....R}, the turbulent diffusivity
closely follows the square of the turbulent velocity profile. From this
comparison, we can infer a correlation time $\tau$ of the turbulence of
${\approx 3\Myr}$. We measure $\eta_{\rm t}$ to be around $2\kpc\kms$, with only
a weak dependence on the applied supernova rate. Improving over previous
studies, we determine the off-diagonals of ${\tilde{\boldsymbol\eta}}$ and find
a positive value of $\delta_z\approx 0.5\kpc\kms$. According to the dispersion
relation in \cite{2005AN....326..787B}, a positive value for $\delta_z$ will,
in principle, allow for a contribution of this term to the dynamo effect.
Turbulent pumping is found to be directed inward and has an amplitude of
$|\gamma_z|\approx 5\kms$. Most strikingly, in our simulations the inward
pumping is balanced by a wind of approximately the same magnitude.
Consequently, the effective vertical transport is drastically reduced,
relieving the dynamo process of a major burden.

\subsection{Field structure and amplification}

In Fig.~\ref{fig:B_of_z} we plot vertical profiles of the exponentially
growing mean radial and azimuthal magnetic field for model H4. We find the
field to be largely confined to the inner disk with a radial pitch angle of
$-10\degr$. Although this is somewhat smaller than the observed values of up to 
$-30\degr$, this value is considerably higher than would be expected from
classical $\alpha\Omega$-dynamos. While the predominant symmetry with respect 
to the midplane is found to be quadrupolar, this mode is interrupted by field
reversals showing dipolar symmetry. This distinct behaviour was successfully
reproduced in a 1D toy model, where the frequency of the periodic field
reversal depended critically on the interplay of the diamagnetic pumping and
the mean vertical velocity profile.

\begin{figure}\begin{center}
  \includegraphics[width=0.9\columnwidth]{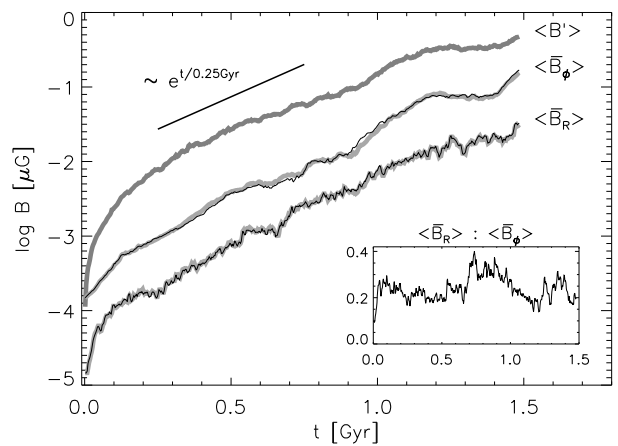}
  \caption{Evolution of the turbulent and regular magnetic field for model
    H4. For $\rms{\bar{B}_R}$ and $\rms{\bar{B}_{\phi}}$ we show the results
    from the direct simulation (grey lines) together with the reconstruction
    from $\EMF(z,t)$ (black lines).}
  \label{fig:B_of_t}
\end{center}\end{figure}

To study the temporal evolution of the arising fields we introduce vertically
integrated rms-values $\rms{\bar{B}_R}$ and $\rms{\bar{B}_{\phi}}$. The 
exponential growth of the field is illustrated in Fig.~\ref{fig:B_of_t}. 
The e-folding time is on the order of $250\Myr$ and varies with the reversals.

In our simulations, the turbulent component dominates over the regular by a
factor of $2-3$. In particular, we find values $\rms{\bar{B}}:\rms{B'}$ of
$0.52\,(\pm0.02)$ for model Q4, $0.40\,(\pm0.03)$ for model H4, and
$0.31\,(\pm0.01)$ for model F4. This trend is consistent with observations of
strong regular fields in the inter-arm regions of spiral galaxies
\citep{2007A&A...470..539B}. Furthermore, from IR-based star formation rates, 
\cite{2007arXiv0712.4175C} observes a correlation 
$\log\ (B_{\rm reg}:B_{\rm tur}) 
\propto -0.32\,(\pm 0.01) \log {\rm SFR}$.  
From our values cited above we find a somewhat steeper slope of
$-0.38\,(\pm0.01)$ with respect to the SN-rate.

\begin{figure}\begin{center}
  \includegraphics[width=1.00\columnwidth]{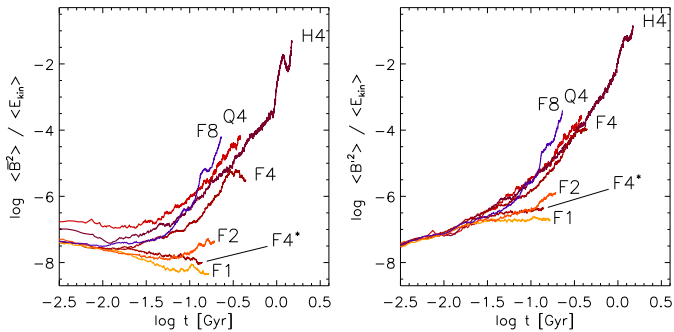}
  \caption{Comparison of the time evolution of the regular and fluctuating
    magnetic field strength over kinetic energy for the various setups
    (cf. Tab.~\ref{tab:models}). Model F4$^{\star}$ is identical to model F4
    with Coriolis forces disabled and demonstrates that no field amplification
    is obtained in the case of Cartesian shear.}
  \label{fig:B_of_t_comp}
\end{center}\end{figure}

From the left panel of Fig.~\ref{fig:B_of_t_comp}, where we plot the time
evolution of the regular and fluctuating components of the various models, we
see that even the absolute value of the mean field $\rms{\bar{B}}$ increases
with decreasing SN-rate (models F4, H4, and Q4). This is consistent with the
trend in the turbulent diffusivity $\eta_{\rm t}$, which has values of $2.3$,
$1.7$, and $1.4\kpc\kms$, respectively.

For the range of parameters studied, we do not observe a significant dependence
of the growth rate on the supernova frequency $\sigma$ but only on the
rotation rate $\Omega$. For the models F1--F8, we find e-folding times of 
$\approx\!500$, $140$, $102$, and $54\Myr$ for the amplification of $\rms{B'}$.
With exception of model F1, which directly corresponds to the parameters used in
\cite{1999ApJ...514L..99K}, we find exponentially growing regular fields 
$\rms{\bar{B}}$ with e-folding times of $147$, $102$, and $52\Myr$ 
for model F2, F4, and F8, respectively. For model F1 the regular field decays 
at $\approx\!500\Myr$. The listed values have been obtained for a time frame of
about $100\Myr$ after the turbulence reaches a steady state. Due to the
different time base these values are not directly comparable to the long-term
growth rate of model H4, where field reversals become important. The initial 
amplification timescale for the models H4 and Q4 is $90\Myr$.

\section{Discussion \& Conclusions}\label{discussion}

We have performed direct simulations of a supernova-driven galactic dynamo and
find the rotation frequency to be the critical parameter allowing the dynamo
to operate.  For our setup, it turns out that $\Omega > 25\kms\kpc^{-1}$ is
necessary, which coincides with the prediction of \cite{1994A&A...286...72S}.
Nevertheless, this value may still be dependent on the assumed gas density and
the gravitational potential.

While the e-folding time of the amplification mechanism scales with the
rotation period, within the range of parameters studied, it only shows a minor
dependence on the supernova rate. With $\tau_e \simeq 250\Myr$, the overall
growth time for our model H4 is comparable to the value obtained for the
cosmic ray driven dynamo \citep{2006AN....327..469H} and about four times
larger than the expected growth time for the magneto-rotational instability.

For similar models without shear \citep{2008AN....329..619G}, we found no
amplification of the mean field. Further simulations will have to show whether
the differential rotation is indeed essential for the SN-dynamo to work, or if
it was only that the critical values for dynamo action were not achieved in
the rigid rotating case given the assumed density profile and gravitational
potential. An estimation based on dynamo numbers leads to slightly subcritical
values for mean-field dynamo action in the case of rigid rotation. 

Although the effects of numerical resolution have to be investigated by
further simulation runs, we are confident that the main features are indeed
robust with respect to the numerical modelling. The good agreement of the
direct simulations with the 1D toy model, as well as the absence of field
amplification in the case without Coriolis forces, provide further evidence in
support of the classical picture of cyclonic turbulence. That is, the Coriolis
force plays the central role in giving the expanding remnants a characteristic
handedness. The mean vertical velocity, together with the turbulent pumping
term, add a new timescale to the system, which is probably responsible for the
fast growth. This was already observed in mean-field models with a prescribed
wind in the mean velocity \citep{1995A&A...297...77E}. 

The relative strength of the regular field compared to the turbulent field is
determined by the SN rate.  The resulting slope of $-0.38\,(\pm 0.01)$ is
similar to the observed one derived from comparisons of polarised over total
intensity of radio synchrotron emission with SFRs \citep{2007arXiv0712.4175C}.
The strong radio-FIR correlation also supports a strong dependence of the
total magnetic field strength with the SFR. Therefore, the saturation level of
the dynamo with respect to the SN-rate should be compared in the future.

Since we do not observe evidence for catastrophic quenching, our results
suggest that the combined action of the diamagnetic pumping and the wind might
cause an outward helicity flux while retaining the mean field. This poses a
natural solution to the quenching catastrophe and will be the subject of
further investigations at higher magnetic Reynolds numbers.


\begin{acknowledgements}
  We gratefully acknowledge Simon Glover and the anonymous referee for helpful
  comments on the manuscript. This project was supported by the Deutsche
  Forschungsgemeinschaft (DFG) under grant Zi-717/2-2.
\end{acknowledgements}


\end{document}